SUBJECT AREAS

Correspondence and requests for materials

# Calibration of multi-layered probes with low/high magnetic moments

Vishal Panchal[1], Héctor Corte-León[1,2], Boris Gribkov[1,3], Luis Alfredo Rodriguez[4,5], Etienne Snoeck[4], Alessandra Manzin[6], Enrico Simonetto[6,7], Silvia Vock[8], Volker Neu[8], and Olga Kazakova[1*]

[1]National Physical Laboratory, Teddington, Hampton Road, TW11 0LW, United Kingdom

[2]Royal Holloway, University of London, Egham Hill, Egham TW20 0EX, United Kingdom

[3]Institute for Physics of Microstructures RAS, Nizhny Novgorod, 603950, Russia

[4]CEMES-CNRS, 29 Rue Jeanne Marvig, B.P. 94347, F-31055, Toulouse, France

[5]Department of Physics, Universidad del Valle, A. A. 25360, Cali, Colombia

[6]Istituto Nazionale di Ricerca Metrologica, I-10135, Torino, Italy

[7]Politecnico di Torino, I-10129, Torino, Italy

[8]Leibniz Institute for Solid State and Materials Research, D-01069 Dresden, Germany

*olga.kazakova@npl.co.uk

**We present a comprehensive method for visualisation and quantification of the magnetic stray field of magnetic force microscopy (MFM) probes, applied to the particular case of custom-made multi-layered probes with controllable high/low magnetic moment states. The probes consist of two decoupled magnetic layers separated by a non-magnetic interlayer, which results in four stable magnetic states: ±*ferromagnetic* (FM) and ±*antiferromagnetic* (A-FM). Direct visualisation of the stray field surrounding the probe apex using electron holography convincingly demonstrates a striking difference in the spatial distribution and strength of the magnetic flux in FM and A-FM states. *In situ* MFM studies of reference samples are used to determine the probe switching fields and spatial resolution. Furthermore, quantitative values of the probe magnetic moments are obtained by determining their real space tip transfer function (RSTTF). We also map the local Hall voltage in graphene Hall nanosensors induced by the probes in different states. The measured transport properties of nanosensors and RSTTF outcomes are introduced as an input in a numerical model of Hall devices to verify the probe magnetic moments. The modelling results fully match the experimental measurements, outlining an all-inclusive method for the calibration of complex magnetic probes with a controllable low/high magnetic moment.**

Magnetic force microscopy (MFM) is a specific mode of scanning probe microscopy, which allows the acquisition of magnetisation distribution on a sample surface with spatial resolution down to a few tens of nanometres[1,2]. Despite its wide-spread use, MFM has several shortcomings. For example, in standard MFM phase imaging, the measurements do not reveal *quantitative* information about the sample stray fields, but merely *qualitative* information about the second-order derivative of the magnetic stray field interaction between the sample and probe. To overcome this obstacle, the magnetic probe has to be calibrated using a well-known reference sample, as, for example, was proposed in Refs. 3–7. Quantitative measurements require a precise characterisation of the probe's properties and a subsequent 'subtraction' of the probe-sample coupling contribution from the measured MFM data[3,8]. Another shortcoming of standard MFM is uncontrollable switching of magnetisation in soft magnetic structures due to strong interaction with the relatively hard magnetic coating of MFM probes[9,10] or, vice versa[9–13]. In this situation, multi-layered (ML)-MFM probes[9–13] that consist of two ferromagnetic layers separated by a non-magnetic interlayer are advantageous due to their ability to be controllably switched between a high moment *ferromagnetic* state (FM: ↓↓, with the layers magnetised in the same direction), and a low moment *antiferromagnetic* state (A-FM: ↓↑, with the layers magnetised in the opposite directions resulting in a closed magnetic field flux around the apex of the probe). This unique property of ML-MFM probes makes them ideal for imaging magnetic structures with a wide range of coercivity and demagnetising fields. However, the interpretation of MFM phase images obtained using ML-MFM probes still requires detailed knowledge of the probe magnetisation and stray field profile.

One option to study the magnetic field geometry of ML-MFM probes is using electron beam techniques that can reveal magnetic domains within and stray field outside ferromagnetic samples. Modes such as Lorentz microscopy[2], differential phase contrast[14] and electron holography (EH)[15,16] have already been employed for imaging magnetic domains, with the latter being particularly useful as it can provide a two-dimensional (2D) map of the projected magnetic flux distribution around the apex of the probe. Although EH is an extremely useful tool, the reconstructed phase images, being a projection of the flux, can often be difficult to interpret due to spurious interactions of the electron beam with magnetic fields emanating from other magnetic objects or electrical charges present in the vicinity of the studied nano-object[17]. Thus, accurate interpretation of EH images require detailed knowledge of the inherent magnetic structure of MFM probes, together with magnetic simulations[18].

Another option is to use Hall sensors (*e.g.*, made of graphene), which have the ability to carry large amount of current and the surface carriers can also be doped to a low carrier density, thus providing high sensitivity to magnetic fields[19]. As a result, sub-micrometre graphene sensors demonstrated a very good ability to detect relatively small magnetic fields with high spatial resolution[20–22]. Characterisation of MFM probes by Hall sensors is generally achieved using the scanning gate microscopy (SGM) technique with frequency-modulated Kelvin probe force microscopy (FM-KPFM) feedback to eliminate any undesirable probe-sample electrostatic effects[21,22]. However, reconstructing the stray field from the interaction between the probe and the Hall sensor is a mathematically complex and computationally time consuming.

The third possible option is define the real space tip transfer function (RSTTF) by means of a quantitative evaluation of the MFM signal taken from a reference sample with very well-known magnetic properties and then to derive magnetic properties of MFM probes[3,8,23,24]. However, although this technique can be used to predict the response of the probe given a magnetic charge map[18], it applies a number of assumptions about probe – sample interaction and cannot be used in cases where the presence of the probe modifies the properties of the sample.

In this paper, we use a comprehensive set of all experimental and modelling methods mentioned above to provide an input to a 2D finite element numerical model, which is used to predict the voltage response of a graphene Hall sensor. This prediction is further validated by the experimental mSGM mapping of the Hall response, providing a comprehensive method for calibration of magnetic probe stray fields. The method is applied to custom-made ML-MFM probes with different thickness and in different magnetic states designed to image samples with both soft and hard magnetisation areas. A commercial single layer coated PPP-MFMR probe (Nanosensors)[25] was also used for comparison. We first define the switching fields required to re-magnetise the ML-MFM probe from the FM to A-FM state by performing MFM phase imaging on a floppy disk sample. We establish a better spatial resolution as achieved by probes in the A-FM state than in the FM state, *i.e.*, about 2 times smaller features can be resolved as proved by imaging a high density hard disk drive (HDD) sample and applying the 20%-80% Edge Spread Function defined in Standards on Lateral Resolution[26]. Then, we directly image the magnetic stray field of the ML-MFM probes in the FM and A-FM states using an *in-situ* EH technique. Furthermore, we derive the real space tip transfer function (RSTTF), *i.e.,* the stray field derivative profile $dH_z/dz(x,y)$ below the apex of the ML-MFM probes, from quantitative MFM (qMFM) measurements of a [Co(0.4 nm)/Pt(0.9 nm)]$_{100}$ multi-layered reference sample. This RSTTF is afterwards integrated to obtain the 2D stray field contours at the same distance (55 nm) below the probe apex and fitted by a double layer dipole model to derive a simplified but still accurate and quantitative description of the ML-MFM probes in various magnetisation states. Finally, we perform Hall voltage mapping of 200 nm-wide single layer epitaxial graphene Hall sensors using magnetic SGM with FM-KPFM feedback for different magnetic states and orientations of the magnetisation for both commercial and ML-MFM probes. These maps were compared against a numerical model that uses the double layer point dipole approximation of the probes to calculate the electric potential in the Hall sensor and, thus, to reconstruct the Hall voltage maps, including the effects of localised magnetic fields and capacitive coupling that can arise from potential differences between the probe and sensor[20,27].

With these experimental and modelling techniques, we demonstrate that the ML-MFM probes can be reliably and controllably switched to any one of the four ±FM and ±A-FM states by applying *ad hoc* magnetic field pulses. The magnitude of the stray field emanating from the apex of the probe depends on the multi-layer coating thickness and mutual orientation of the layers as verified by EH imaging and Hall voltage maps, and is directly quantified by integrating the RSTTF. In the FM state, the stray field is transversely originated from the apex of the probe, whereas in the A-FM state a closed magnetic field flux is observed. These features should be taken into account when analysing MFM phase images taken with ML-MFM probes.

## Results

### Controllable switching of ML-MFM probes magnetic states

In order to establish the switching fields and sensitivity of the four different magnetic states of the ML-MFM probes, a reference floppy disk sample was scanned in the MFM phase imaging mode. To control the magnetic configuration of ML-MFM probe, we applied a pulse of out-of-plane magnetic field $B_\perp$ (*i.e.*, parallel to the vertical axis of the probe). Figure 1a shows MFM images obtained in four different configurations (±FM and ±A-FM) for the thin ML-MFM probe. The ±FM states were achieved by applying a 10 ms pulse of $B_\perp = \pm 20$ mT, whereas the ±A-FM states were achieved by applying consequent pulses of $B_\perp \approx \mp 20$ mT and $\pm 13$ mT (Fig. 1b). The line profiles taken along the same region of the floppy disk sample for the −FM and −A-FM states clearly show that −FM state exhibits approximately double the phase change compared to the −A-FM state (Fig. 1c). Comparison of the line profiles for the ±FM (Fig. 1d) and ±A-FM (Fig. 1e) states demonstrates the expected inversion of the MFM phase change, which indicates successful reversal of the probe magnetisation and its stability during the scanning. Additionally, imaging a HDD sample with bit size of 30 nm revealed average lateral spatial resolution of thin ML-MFM probe in FM and A-FM states to be 21.4±4.1 nm and 12.6±2.2 nm, respectively (see Supplementary Fig. S1 and Table S1).

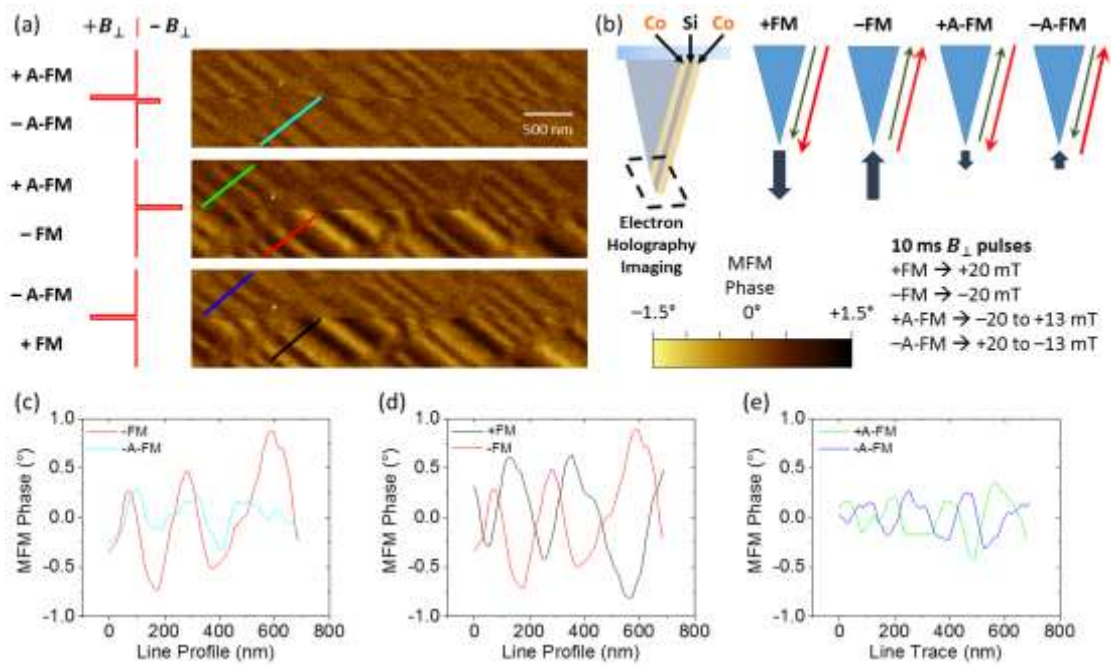

**Figure 1. Switching of ML-MFM probes magnetic states.** (a) MFM phase images of a reference floppy disk sample obtained with a thin ML-MFM probe in four different configurations (±FM and ±A-FM). Each of the three MFM phase images were obtained by continuously scanning from top to bottom, while applying pulses of the out-of-plane magnetic field ($B_\perp$, as represented by the red line schematics on the left) to switch the magnetic states of the probe. (b) Schematic representation of the four different ML-MFM probe configurations. Line profiles for (c) −FM and −A-FM, (d) +FM and −FM, and (d) +A-FM and −A-FM states, obtained along the lines of the corresponding colour in (a).

**Scanning electron microscopy and electron holography imaging**

Figure 2a shows a scanning electron microscopy (SEM) image with schematic overlay of the direction of the Co/Si/Co multi-layer deposition on a Si probe. For the thin and thick ML-MFM probes, the final curvature radius is ~20 nm and ~35 nm, respectively. Using the pulsed field sequences (identified in the previous section), the ML-MFM probes were magnetised in FM and A-FM states and the stray field geometry for each of the states was imaged by *in-situ* magnetic field EH in order to investigate the different MFM response between FM and A-FM states observed in Fig. 1. Figure 2b displays EH images for thick ML-MFM probe in ±FM and +A-FM states. The colour images represent the magnetic phase shift produced around the probe apex, while the black and white images illustrate the respective magnetic flux lines of the stray field (see Method section). Imaging the magnetic phase shift in vicinity of the apex of the thick ML-MFM probe, we found that the magnetic flux lines in the ±FM states have a similar geometry to those reported for commercial uniformly coated MFM probes[15,16,28], where the field emerges/enters almost perpendicular to the MFM probe surface. This configuration allows for strong interaction with the sample's magnetisation. Moreover, we clearly see that the phase shift gradients for the +FM/−FM states are oppositely oriented, signifying a reversal of the magnetic flux lines direction. By contrast, in the A-FM state the magnetic flux lines curl around the apex due to the magnetic coupling between the north and south poles of the ferromagnetic Co layers. In this curling phase, the stray field is parallel to the sample surface. Thus, the A-FM state is significantly less invasive, see also Fig. 1, allowing for studies of magnetically sensitive samples (*e.g.*, materials with low coercivity, devices with weakly pinned domain walls, skyrmions, etc.[10]). Furthermore, MFM probes with a horizontal magnetic field component could potentially be exploited in orientation-sensitive MFM phase imaging[29] or vector MFM. EH images for the thin MFM-probe in the ±FM states, exhibiting a similar behaviour to that of the thick ML-MFM probe are presented in Supplementary Fig. S3.

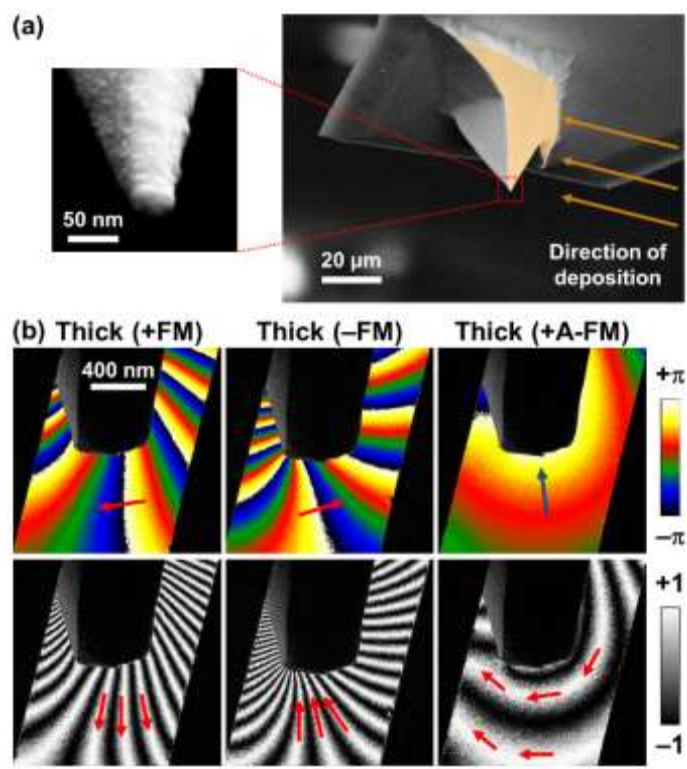

**Figure 2. Scanning electron micrograph and electron holography images of ML-MFM probes.** (a) SEM images illustrating the deposition of multi-layer coating on a Si probe, final curvature radius is ~20 nm and ~35 nm for the thin and thick ML-MFM probes, respectively. (b) EH images taken near the apex of the thick ML-MFM probe. Colour images correspond to the magnetic phase shift, while black and white images represent the configuration of the magnetic flux due to the stray field. Arrows in the phase shift and magnetic flux images indicate the direction of the phase shift gradient and magnetic flux, respectively.

**Quantifying the RSTTF of ML-MFM probes**

Figure 3a displays the MFM measurements conducted on the Co/Pt multi-layered reference sample with the thin ML-MFM probe in its four different magnetisation states. Contrast reduction between FM and A-FM states and contrast inversion between up (−) and down (+) magnetisation states are clearly visible. Exemplary line profiles taken at the exact same position according to the topography channel are compared in Fig. 3b, corrected only for an overall phase shift of the individual images. Inverting the contrast for the +A-FM and +FM states and comparing them with −A-FM and −FM states result in a perfect quantitative agreement between the probe's response in the − and + configurations, which demonstrates the successful reversal of the probe magnetisation and its stability during scanning (Fig. 3c). Interestingly, in the A-FM state the MFM profiles are not only varied in amplitude by a factor of two, as expected, but also slightly shifted by about 30 nm along the *x*-axis. We impute this shift to an effect of the magnetic apex (where the stray field reaches its peak value) not necessarily being at the same position as the physical probe apex. A plausible explanation is the presence of additional in-plane components of the magnetisation state, when the ML-MFM probe is in its A-FM flux closure configuration, which will shift the measured MFM-profiles above the labyrinth domain pattern of the reference sample. The suggestion of an additional in-plane component is consistent with the magnetic flux distribution for the +A-FM state shown in Fig. 2b.

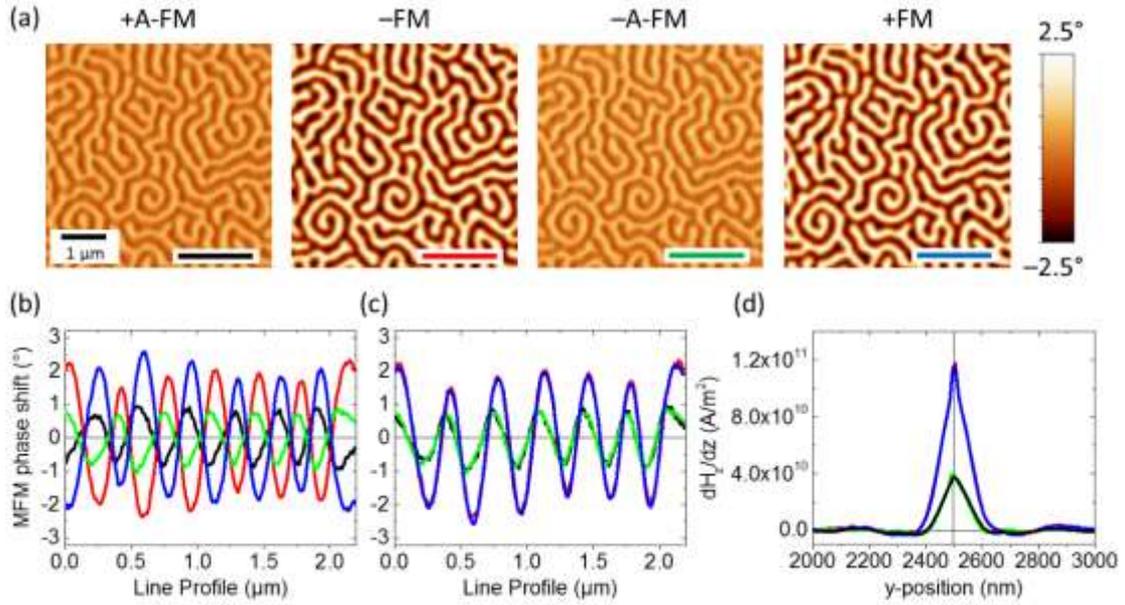

**Figure 3. RSTTF of thin ML-MFM probe.** (a) MFM phase shift measurements on the Co/Pt reference sample using the thin ML-MFM probe magnetised in the (left to right) +A-FM, −FM, −A-FM and +FM states, respectively. (b) MFM profiles obtained along the lines of the corresponding colour in (a). (c) Same MFM profiles as reported in (b), where the +A-FM and +FM profiles are inverted to be compared with the −A-FM and −FM states. (d) Real space tip transfer function of the four different magnetisation states (one-dimensional cut along the image y-direction, which is perpendicular to the cantilever length).

A cross section of the RSTTF of the ML-MFM probe along the *y*-axis in its various magnetisation states is reported in Fig. 3d. The stray field derivative profiles at a distance of 55 nm below the physical probe apex are given in positive values, independently of the polarity of the probe. As expected, the profiles of − and + magnetisation configurations lie on top of each other and the profiles in the A-FM states are strongly reduced over those of the FM state. The RSTTF is a true quantitative characterisation of the ML-MFM probe and can be used to quantitatively analyse MFM measurements of unknown samples. In the present study, RSTTF is used to calculate the probe's stray field profile $H_z(x,y)$ by direct integration to obtain a simplified description of the ML-MFM probe with a two-layer point dipole model and to reconstruct the expected SGM signal (see next section). The parameters of this two-layer dipole model are determined to simultaneously give a good description of four $H_z$-profiles, namely $H_z^{(y)}$ and $H_z^{(y)}$ in the FM- and the A-FM state. Considering only the symmetrical one-dimensional stray field profiles along the *y*-direction (perpendicular to the cantilever length, Fig. 3d), the thin ML-MFM probe can be rather well described by two point dipoles with different magnetic moment z-components (vertical) $m_z^{(1)} = 7.0 \times 10^{-17}$ A·m² and $m_z^{(2)} = 5.0 \times 10^{-17}$ A·m². They are positioned inside the probe at a distance of about 99 nm from the probe apex and are separated by 25 nm along the x-directions parallel to the sample plane (Fig. 4). This approximation is valid for both FM and A-FM states, with the two dipoles pointing along the same or opposite directions, respectively. Analysing and fitting stray field profiles along the x-direction, a slight asymmetry in the profiles is observed, which requires a rotation of the dipole moments by about 8° (Fig. 4a). The additional x-components (horizontal) of the magnetic moment have values of $m_x^{(1)} = 1.0 \times 10^{-17}$ A·m² and $m_x^{(2)} = 0.7 \times 10^{-17}$ A·m². The complete parameter set for both thin and thick ML-MFM probes in the FM and A-FM states is summarised in Table 1. The qMFM measurements and analysis of the thick ML-MFM probe are given in the Supplementary Fig. S2.

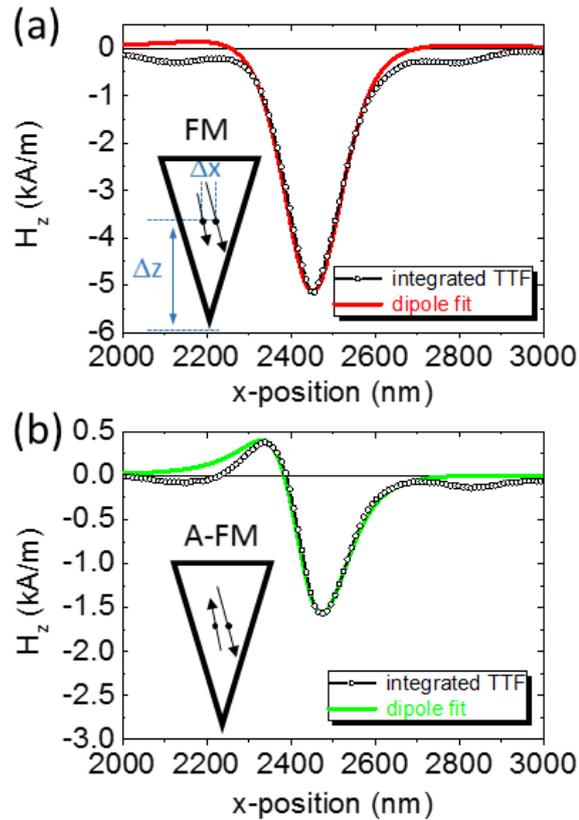

**Figure 4. Stray field profiles obtained from integrated RSTTF.** Stray field profiles for the thin ML-MFM probe in the (a) FM and (b) A-FM states, obtained 55 nm below the probe apex. Δx and Δz are horizontal and vertical displacement of the two dipoles, respectively. The profiles are fitted with the two-layer point dipole parameters listed in Table 1.

**Table 1.** $z$- and $x$-components ($m_z^{(i)}$, $m_x^{(i)}$) of the dipole moments of each individual layer (i=1,2) within thin and thick ML-MFM probes and their vertical displacement ($\Delta z$). The first five parameters are a result of the fitting procedure, while the horizontal displacement of the two dipoles ($\Delta x$) is given by the layer thicknesses.

| MFM Probe | State | $m_z^{(1)}$ (A·m²) | $m_z^{(2)}$ (A·m²) | $m_x^{(1)}$ (A·m²) | $m_x^{(2)}$ (A·m²) | $\Delta z$ (nm) | $\Delta x$ (nm) |
|---|---|---|---|---|---|---|---|
| Thin ML | FM | −7.0×10⁻¹⁷ | −5.0×10⁻¹⁷ | 1.0×10⁻¹⁷ | 0.7×10⁻¹⁷ | 99 | 25 |
| Thin ML | A-FM | −7.3×10⁻¹⁷ | +5.0×10⁻¹⁷ | 1.0×10⁻¹⁷ | 0.7×10⁻¹⁷ | 99 | 25 |
| Thick ML | FM | −13.0×10⁻¹⁷ | −9.0×10⁻¹⁷ | 1.0×10⁻¹⁷ | 0.6×10⁻¹⁷ | 113 | 40 |
| Thick ML | A-FM | −13.0×10⁻¹⁷ | +8.0×10⁻¹⁷ | 1.0×10⁻¹⁷ | 0.6×10⁻¹⁷ | 113 | 40 |

Apart from small deviations in the absolute moment values, the switching from the FM to the A-FM state essentially occurs via the reversal of the $z$-component of the layer with smaller dipole moment. This simplified description of the true 2D stray field characteristic allows for an easy estimation of the stray fields at various distances below the probe apex and is used for SGM calculations.

**Magnetic scanning gate microscopy**

Figure 5a shows the experimental mSGM measurement setup for mapping the local Hall voltage ($V_H$) with 15 µA bias current applied across the single layer graphene Hall sensor and peak-to-peak probe oscillation amplitude ($A_{osc}$) of 88 nm (see Methods for further details). The minimum distance of the probe from the sensor plane is zero and the probe in-plane projection is orientated at ∼10° from the vertical arm of the Hall cross (Fig. 5b). Figure 6 shows the experimental mSGM maps for commercial as well as thin and thick ML-MFM probes in FM and A-FM (only for ML-MFM probes) configurations. The magnitude of the $V_H$ response depends on the probe-sample vertical separation, the oscillation amplitude and magnetic moment of the MFM probe and the bias current applied to the device. The polarity of $V_H$ depends on the directions of the applied current and on the probe magnetisation orientation[21,22]. Thus, in the present dataset, parameters for the probe-sample separation, oscillation amplitude, bias current and electrical connections have been kept the same throughout the experiment. First, a stable reference map of $V_H$ was established using a commercial probe in +/− orientation of the stray field with the peak signal of $V_H$ ~+1.7/−1.9 µV, respectively (Fig. 6a and Table 2).

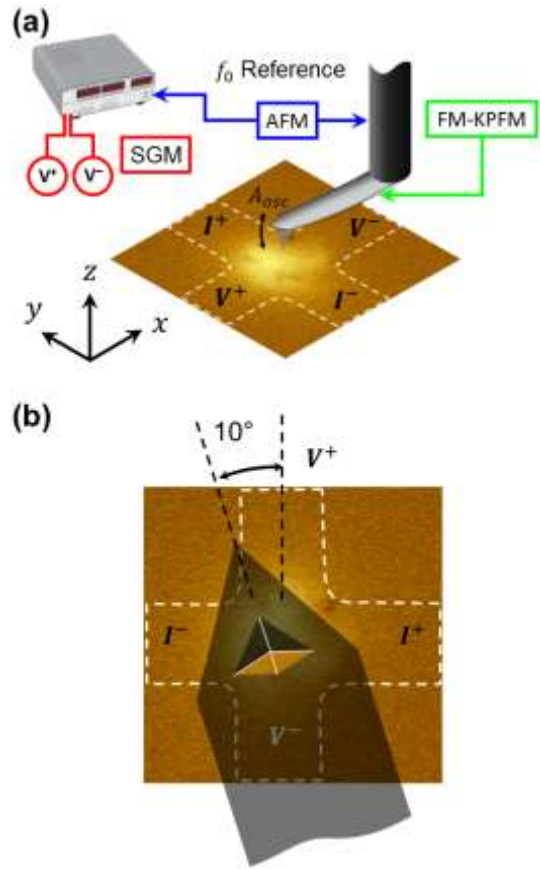

**Figure 5. Experimental magnetic SGM setup and graphene Hall cross.** (a) Surface potential image of the 200 nm Hall cross made of single layer graphene at zero bias, obtained with frequency-amplitude Kelvin probe force microscopy (FM-KPFM). (b) Schematics of the single-pass scanning gate microscopy (SGM) mode with FM-KPFM feedback to eliminate the undesirable electrostatic interaction between the MFM probe and Hall sensor. (c) Schematic representation showing the 10° in-plane rotation of the cantilever with respect to the vertical arm of the Hall cross.

In the +FM state of the thin and thick ML-MFM probes, peak $V_H$ values of ~+3.0 µV (Fig. 6b) and ~+5.1 µV (Fig. 6c), respectively, was observed within the Hall cross region. The line profiles in Fig. 6d show a clear bell type response. The increase in $V_H$ response is due to the increment in the Co layer thickness, which directly leads to an increase in the probe's stray magnetic field. We can assume that possible formation of multi-domain states in the thick probe does not significantly affect the magnetisation at the apex of the probe and thus an overall increase in the probe magnetic moment can be detected. Similar Hall images and thickness dependence were also observed for commercial single layer probes from other manufactures[21,30]. Similarly to the commercial probe, the remagnetisation of ML-MFM probes to the −FM state leads to the change in the $V_H$ polarity, *i.e.*, bell shape response with the negative peaks (Figs. 6a-6c and 6e). The relative changes in the peak $V_H$ values were generally consistent for all the analysed probes in their respective −FM states (−1.9 µV, −3.1 µV and −6.4 µV for commercial, thin and thick ML-MFM probes, respectively) (Fig. 6e and Table 2). Moreover, the radial symmetry of the Hall response for the ±FM states suggests that the magnetisation is mainly aligned along the *z*-axis of the probe, which is in good agreement with electron holography images (Fig. 2b) and the dipole approximation from the RSTTF (Table 1).

**Table 2.** Summary of peak $V_H$ values extracted from Fig. 6 for commercial, thin and thick ML-MFM probes.

| Magnetisation State | Hall voltage [µV] | | |
|---|---|---|---|
| | Commercial 30 nm CoCr | Thin ML-MFM 15/10/15 nm Co/Si/Co | Thick ML-MFM 30/10/30 nm Co/Si/Co |
| +FM | +1.7 | +3.0 | +5.1 |
| −FM | −1.9 | −3.1 | −6.4 |
| +A-FM | N/A | +0.5 | +2.6 |
| −A-FM | N/A | −0.75 | −2.8 |

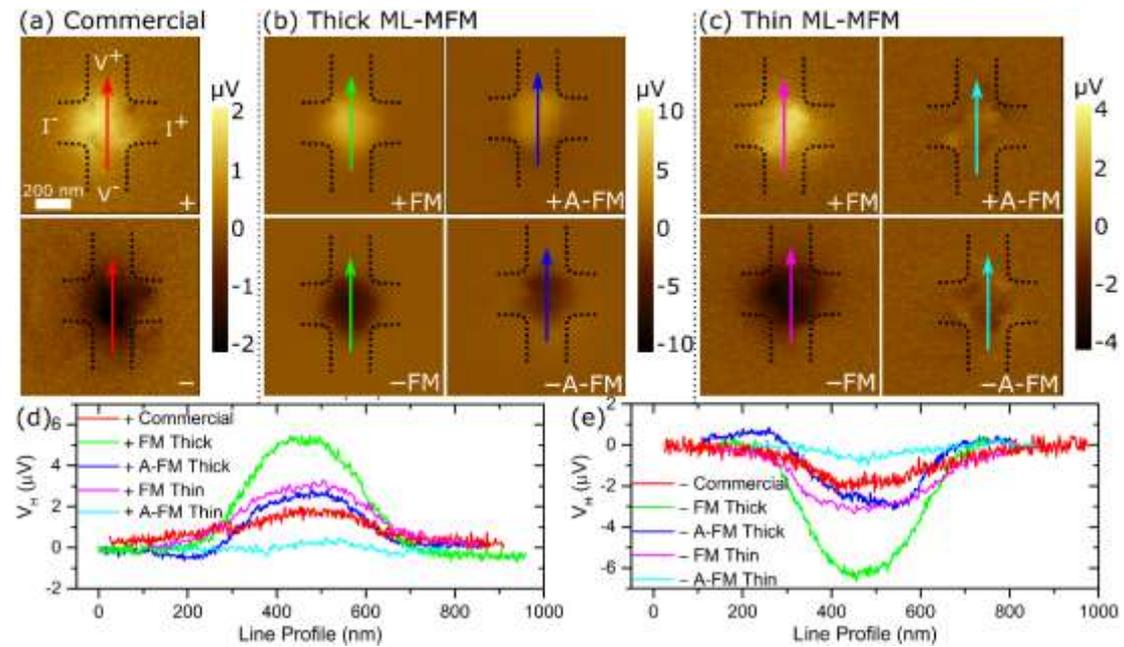

**Figure 6. Experimental magnetic SGM maps of local Hall voltage.** Magnetic SGM images for (a) commercial Nanosensors MFM probe, (b) thick and (c) thin ML-MFM probes in the FM (↓↓) and A-FM states (↓↑). Top/bottom rows of images in (a), (b) and (c) are for probes magnetised +/−, respectively. Black dashed lines depict the Hall cross

borders. The electrical connections shown in (a) are the same for all other images. (d) and (e) are line profiles of the Hall voltage across the lines of the same colour indicated in (a)-(c) for the probes magnetised in + and − direction, respectively.

For the *thick* ML-MFM probe, the $V_H$ peak in the ±A-FM state becomes ~2 times smaller than in the ±FM state, which signifies a noticeable decrease in the out-of-plane magnetic field coming from the probe apex (Figs. 6d, 6e and Table 2). It is noteworthy that for the thin ML-MFM probes, the previous comparison (Fig. 1 and Supplementary Fig. S1) between FM and A-FM states yielded ~2- and ~4-fold decrease in MFM phase signal for floppy and HDD sample, respectively. This discrepancy is likely related to material properties of the floppy disk and HDD, and also their magnetisation direction (*i.e.*, parallel and perpendicular recording, respectively). In the current experiment, an even larger decrease (~5-fold) is observed for the *thin* ML-MFM probe transitioning from ±FM to ±A-FM, which indicates the pronounced reduction in the stray magnetic field. As a consequence of the generally lower magnetic response from the A-FM state of the *thin* ML-MFM probe, the maps also reveal minor electrostatic signal due to imperfect KPFM compensation, seen as the dark and bright contrasts at the corners of the Hall cross (Fig. 6c, right column)[21,22]. Closer inspection of the maps from commercial and thick probes also shows small parasitic electrostatic contribution, however in these cases, the significantly larger magnetic response masks the weaker electrostatic signal. Regardless, the lower $V_H$ response from A-FM states is consistent with the dipole approximation (Table 1) and the EH images obtained on thick ML-MFM probe (Figs. 2b), where the magnetic flux lines are wider spaced and have noticeably different geometrical profiles.

The numerical model illustrated in Methods is applied here to reconstruct SGM images using the two point dipoles approximation obtained through the RSTTF. As demonstrated by the comparison of Figs. 6 and 7, a good agreement between measurements and simulations is found for the maps obtained with the *thick* ML-MFM probe, for both FM and A-FM states. For the last case, if in-plane magnetic moment components were absent (the two Co layers were approximated by two dipoles with anti-parallel magnetisation equal in magnitude), a two-fold symmetry would be obtained, with negligible values at the Hall cross centre and balance between positive and negative value regions. However, the presence of important in-plane magnetic moment components together with unbalance between z-components (RSTTF values in Table 1) results in a Hall voltage map with similar spatial distribution to the one obtained with the probe in the FM state, apart from a change in sign at the bottom part of the Hall cross. This is in agreement with the experimental results, confirming the non-trivial magnetisation arrangement of the two magnetic layers, and thus the reliability of the RSTTF parameterisation.

Regarding the images obtained with the *thin* ML-MFM probe, a reliable numerical reconstruction of the measured maps can be achieved only for the FM state when using the parameters extracted from RSTTF characterisation. A careful observation of the A-FM state experimental map in Fig. 6c reveals that in addition to the weak bell shape Hall voltage response, there are also minor signals at the cross corners of the cross with a quasi-four-fold symmetry. This is due to a non-perfect FM-KPFM compensation, which leads to a non-negligible probe-sample capacitive coupling. Line profiles obtained along the diagonal of the Hall cross for both experimental and simulated SGM maps (see Supplementary Figs. S4 and S5, respectively) clearly show the electrostatic contribution in the thin ML-MFM probe in A-FM state. To reconstruct the experimental results in a more reliable way, the simulations are repeated including the capacitive contribution, described by Eqs. (4) and (5) in Methods, by considering a maximum local carrier density variation $\Delta n$ of 0.07% and a characteristic length of the interaction $d = $ 30 nm (Fig. 7b). This electrostatic contribution improves the reconstruction of the map obtained with the thin ML-MFM probe in the FM state, with a voltage signal distribution characterised by a weak eccentricity along one of the main diagonals and two peak regions shifted towards the corners. Thus, in the experimental map obtained with the thin ML-MFM probe, the typical magnetic features of the A-FM state are hidden by the electrostatic probe-sensor interaction, making it difficult to interpret the results if a non-proper model of the probe effects is implemented.

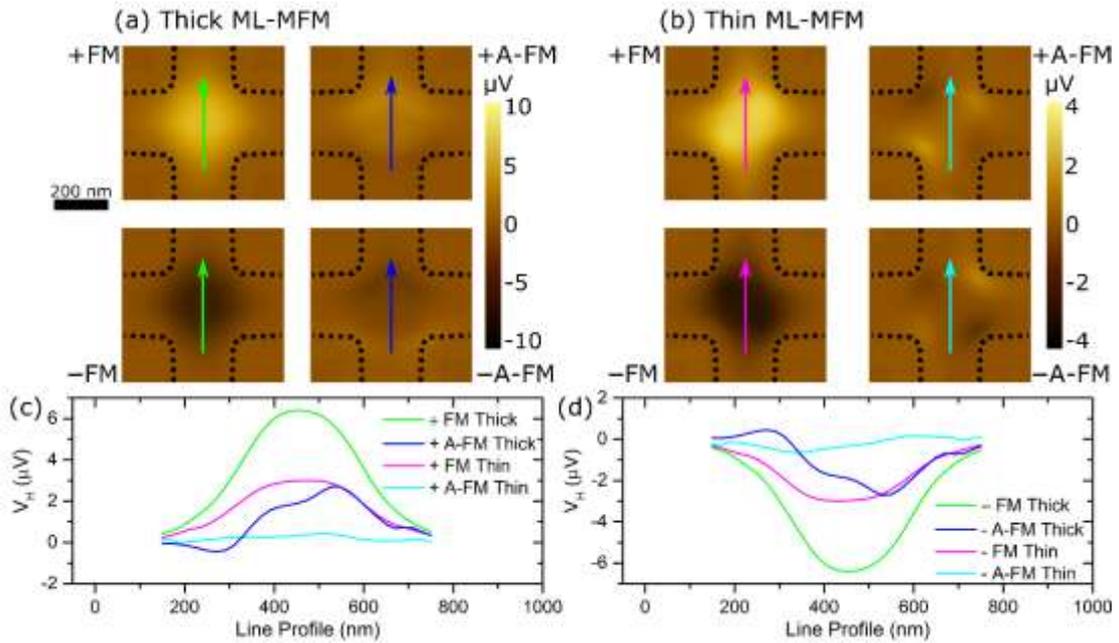

**Figure 7. Simulated magnetic SGM maps of local Hall voltage.** Simulated magnetic SGM images for (a) thick and (b) thin ML-MFM probes in the FM (↓↓) and A-FM states (↓↑). Top/bottom rows of images in (a) and (b) are for probes magnetised +/−, respectively. Black dashed lines depict the Hall cross borders. (c) and (d) are line profiles of the Hall voltage across the lines of the same colour indicated in (a) and (b) for the probes magnetised in + and − direction, respectively.

The complex distribution of the magnetic field near the apex of ML-MFM probes should be taken into account in MFM measurements. Furthermore, the observed thickness dependence of ML-MFM probe behaviour can be fine-tuned to enhance the sensitivity and lateral resolution of MFM as well as to significantly improve the non-invasiveness of MFM phase imaging, which is particularly important for the investigation of soft magnetic materials[10].

**Conclusions**

Using a number of experimental (EH, *in situ* MFM, mSGM) and modelling (RSTTF, finite element model) methods, we have presented a method for visualising and quantifying the magnetic stray field of MFM probes and used it to study custom-made ML-MFM probes of two thicknesses in multiple magnetic states. The combination of the techniques provides a comprehensive picture of the magnetic states of the probes, including switching fields between FM and A-FM states, stability of all states during scanning and spatial distribution of the probe stray field. EH imaging revealed that the stray field below the apex of the ML-MFM probes is generally perpendicular to the sample in the FM state. However, the most notable exception is the thick ML-MFM probe in the A-FM state, where a significant horizontal component was observed. MFM imaging was used to demonstrate that thin ML-MFM probes in the A-FM state provide the highest spatial resolution of ~12 nm, which is almost twice better than for the same probe in the FM state, while maintaining relatively large magnitude of the MFM phase image.

By means of qMFM measurements on a well-defined magnetic sample, the RSTTF was deduced and the stray field profiles of the ML-MFM probes in their various magnetisation states were obtained. Fitting these profiles with a double layer point dipole model, the following features could be quantitatively evaluated: (i) in the FM state, the probes are mainly characterised by a strong and symmetric vertical component of the stray field below the probe apex; (ii) in the A-FM state, this stray field component is strongly reduced and a sizable stray field component along the horizontal direction is induced; (iii) the thick probes possess a stray field ~1.8 times larger than that of the thin probes.

The mSGM technique exploiting well calibrated Hall sensor was demonstrated to be particularly advantageous, allowing direct measurements of the voltage signal proportional to the probe stray field at a certain probe-sample distance. The mSGM technique offered a completely independent approach to quantify the probes' stray fields. The achieved results revealed that both in the FM and A-FM configurations, bell shaped Hall voltage response is observed (similar results were obtained for commercial single layer MFM probes). Predictably, transition from FM to the A-FM configuration leads to ~2-5 times decrease in the Hall voltage response. By using the double layer dipole parameters of the qMFM measurements, the simulated Hall sensor response is in full quantitative agreement with the experimental results, (taking into account the probe-sensor capacitive coupling (notable in the case of the thin probe in the A-FM state).

For the first time, three fully independent experimental approaches used to characterise the magnetic properties of MFM probes – electron holography, qMFM (with RSTTF) and SGM – result in a coherent picture; enabling us to quantify and verify the predicted magnetic moments of the probes with numerical modelling. The custom-made novel ML-MFM probes with low/high moment states can be used for MFM phase imaging with high spatial resolution and sensitivity. The ability to controllably switch their magnetic moment makes them particularly useful for studies of samples with strong/weak magnetisation.

## Methods

**Fabrication of multi-layered probes.** A series of ML-MFM probes were fabricated using magnetron sputtering (AJA International Aurora, ATC-2200) in Ar atmosphere. Commercial Si cantilevers (PPP-FMR, Nanosensors) with typical resonance frequency $f_0$ = 70-80 kHz, force constant = 2-3 Nm$^{-1}$ and curvature radius of ~10 nm were chosen for coating. The coating was deposited on two faces of the pyramidal probe (Fig. 2a) and the ML-MFM probe was comprised of two Co layers separated by a Si interlayer (Fig. 1b). Two coating thicknesses were considered, *i.e.*, Co(30 nm)/Si(10 nm)/Co(30 nm) for thick and Co(15 nm)/Si(10 nm)/Co(15 nm) for thin ML-MFM probe. The film thicknesses were estimated using SEM and material deposition rates measured on a flat surface. The final curvature radii were ~20 nm and ~35 nm for thin and thick ML-MFM probes respectively. For comparison, the curvature radius of commercial MFM probes (PPP-MFMR, Nanosensors[25]) is ~30 nm. Detailed SEM investigations of custom-made ML-MFM probes revealed that the outer magnetic layer is longer, *i.e.*, geometrically closer to the sample's surface, than the inner one, see schematics in Fig. 1b. Furthermore, the orientation of the ML-MFM probe faces was within 2° of being perpendicular to the sample surface during scanning.

**Magnetic force microscopy phase imaging.** The MFM phase imaging of the floppy and hard disk drive sample was performed with the NT-MDT Ntegra Aura scanning probe microscope (SPM). The system was fitted with a home-built coil to apply an out-of-plane magnetic field during scanning. MFM phase imaging was carried out as a two-pass technique using the ML-MFM probes. During the first-pass, the SPM was operated in atomic force microscopy mode to determine the topography. During the second-pass, the topography line (obtained during the first-pass) is retraced while oscillating the probe at $f_0$, maintaining a set distance of 9 nm between the probe and sample, and recording the cantilever phase change resulting from the probe-sample magnetic interactions.

**Electron holography imaging.** EH experiments were carried out in the Hitachi HF3300 (I2TEM-Toulouse) microscope, a TEM specially designed to perform *in-situ* EH experiments with high phase shift sensitivity and spatial resolution lower than 1 nm, thanks to the combination of a high brightness cold field emission gun[31] (of about ~ $10^9$ A/cm²·sr), an image corrector (aplanator B-COR, from CEOS, for correcting off-axial aberrations) and a multi-biprism setup capability[32]. EH is a powerful technique employed to study the local magnetic distribution of ferromagnetic nanostructures by imaging the two-dimensional projection of the magnetic induction inside and outside the specimen. By performing an interferometry experiments, EH retrieves the phase shift of the object electron wave, which is strongly perturbed by the electromagnetic potentials present inside (magnetisation, mean inner potential) and outside (magnetic stray field, electric field) of the nanostructure. Magnetic information of the specimen is obtained from the retrieved phase shift, $\varphi(x, y)$, by solving the following equation:

$$\varphi(x, y) = \int_{-\infty}^{\infty} C_E V(x, y, z) dz - \frac{e}{\hbar} \int_{-\infty}^{\infty} A_z(x, y, z) dz = \varphi_E(x, y) + \varphi_M(x, y)$$

(1)

where $C_E$ is an interaction constant depending on the acceleration voltage of the electron beam (for a 300 kV TEM, $C_E = 6.53 \times 10^6$ rad V$^{-1}$m$^{-1}$), $e$ is the electron charge, $\hbar$ is the reduced Planck constant, $V$ is the electric potential and $A_z$ is the component of the magnetic vector potential, which is perpendicular to the electron trajectory ($z$-axis). In absence of any electric potential, phase shift only provides magnetic information of the sample and it is directly proportional to the magnetic flux, $\Phi(x, y)$, [$\varphi_M(x, y) = (e/\hbar)\Phi(x, y)$][33], so images of the phase shift will directly provide maps of the magnetic flux. Moreover, $\varphi_M(x, y)$ and the projected magnetic induction, $B_{proj}(x, y)$ are related as $\nabla\varphi(x, y) \cdot B_{proj}(x, y) = 0$,[34] so that the direction of the magnetic phase shift gradient is linked with the perpendicular direction of the projected magnetic induction, following the right-hand rules between them $\nabla\varphi(x, y)$, $B_{proj}(x, y)$ and the electron trajectory. In Figure 2b, we represent magnetic flux images of the stray field distribution near the ML-MFM probes' apex for the FM and the A-FM states for the thick ML-MFM probe. In these images, magnetic flux line representation is made where a sinusoidal function is applied on amplified magnetic phase shift images [$\cos(n\varphi(x, y))$, where $n$ is an amplifier factor]. As the pyramidal base of the ML-MFM probes has a size of several microns, the EH setup was tuned to reach the maximum field of view (1.05 μm), with a spatial resolution of 3 nm, using a double-biprism setup. The different magnetisation states where identified using a corrected Lorentz mode and placing the samples in the 'normal stage' of the I2TEM (conventional TEM holder position, where the specimen is located between the pole pieces of the objective lens), after switching off the objective lens. The controlled magnetic field produced by the objective lens pole pieces was used to induce the FM and A-FM states of the ML-MFM probes.

**Quantitative MFM imaging and probe characterisation with a double-layer point probe model.** The qMFM measurements have been performed with a Bruker Icon scanning probe microscope using the Nanoscope V controller. Prior to the measurements, the magnetisation state of the ML probes has been set by bringing them into a similar sequence of perpendicular fields as mentioned in the first section, however this time outside the microscope. The spring constant of the probes' cantilevers was individually measured with thermal tuning and the quality factor of the resonance was determined during the resonance tuning process. For a full quantification of the real space tip transfer function (RSTTF), MFM phase imaging was performed in standard lift mode (total distance between sample surface and probe apex $d_{tot}$ = 55 nm) on a [Co(0.4 nm)/Pt(0.9 nm)]$_{100}$ multi-layered sample with perpendicular magnetic anisotropy (Fig. 3a). This reference sample has well-characterised integral magnetic properties (saturation polarisation Ms = 457 kA/m, perpendicular anisotropy constant $K_u$ = 517 kJ/m$^3$, domain transition width δ = 20 nm), and a pure qualitative MFM image allows calculating the effective surface charge pattern of the sample and thus determining the object which has been imaged. The magnetic behaviour of the probe was quantified by de-convolving the measured MFM image and the effective surface charge pattern by means of a Fourier–based qMFM code implemented in SigMath (for more details see Ref. [23]). The resulting RSTTF presents the probe's stray field derivative profile $dH_z/dz$ (x,y) at the distance $d_{tot}$ below the apex of the probe (Fig. 3d). It is a correct parameter-free characterisation of the magnetic probe, including its 3-dimensional extend and possible non-uniformities in the magnetisation state. Integration of the RSTTF (again in Fourier space) results in the quantitative stray field profile $H_z(x,y)$ (Fig. 4). For offering a more descriptive picture of the various magnetisation states in the ML-MFM probes, cuts of the stray field profile along the cantilever direction ($x$-direction) and perpendicular to it ($y$-direction) (Fig. 5b) have been fitted with a double-layer dipole model (Table 1). In this simplified model, the probe is approximated by two magnetic dipole moments separated along $x$-direction by 25 nm (thin ML-MFM probe) or 40 nm (thick ML-MFM probe) and positioned within the probe in a vertical distance Δ$z$ away from the apex (Fig. 4). According to the layer geometry of the probes, the dipole moments were assumed to have a dominating $z$-component, a possible $x$-component and zero $y$-component. Within one probe, the ratio of $x$ to $z$ component in the two layers was assumed to be approximately equal; its magnitude, however, was allowed to differ. With these constraints, dipole parameters of FM and A-FM state for thin and thick ML-MFM probes were determined. Note that considering only the probe in the FM-state, the fitting procedure is only sensitive to the sum of the dipole moments: $m_z^{(1)} + m_z^{(2)}$ and

$m_x^{(1)}+m_x^{(2)}$, however including the profiles in the A-FM state, where one dipole has a negative $m_z$-orientation, the individual layer contributions can be disentangled.

**Hall sensor fabrication.** 200 nm wide single layer epitaxial graphene Hall sensors on 6$H$-SiC(0001) were fabricated using the procedures reported in our previous publications[35,36]. Magnetotransport measurements performed on the device, using the techniques described in Ref. [37] revealed the electron carrier density, $n_e$ = 5.1×10$^{11}$ cm$^{-2}$, and carrier mobility, $\mu_e$ = 5800 cm$^2$V$^{-1}$s$^{-1}$, at room temperature and in ambient air.

**Magnetic scanning gate microscopy.** The magnetic SGM characterisation of the MFM probes were performed using the NT-MDT Ntegra Aura SPM fitted with a home-build transport measurement stage. Hall characterisation of MFM probes was performed using single-pass SGM mode with FM-KPFM feedback to eliminate the undesirable electrostatic interaction between the MFM probe and Hall sensor (Fig. 5a)[21,22]. During scanning of the Hall cross, the MFM probe is oscillating at the first harmonic of the cantilever resonant frequency ($f_0$), which leads to oscillation of probe's stray magnetic field and therefore oscillation of $V_H$. Using this method, the $V_H$ is recorded at each point of the scan area with a Stanford Research Systems SR830 lock-in amplifier referenced to $f_0$. It should be noted that there is a 10° in-plane rotation of the cantilever relative to the vertical arm of the Hall cross (Fig. 5b). Further details about Hall measurements process can be found in Refs. [21] and [22].

**Numerical modelling of scanning gate microscopy images.** The Hall response to the stray field of the ML-MFM probes is simulated by means of a 2D finite element model[20,27], which enables to calculate the distribution of the electric potential $\phi$ inside the Hall sensor under the assumption of diffusive transport regime. The electron transport is described by the following stationary equation

$$\nabla \cdot \left[ \ddot{\sigma}(\mathbf{r}) \nabla \phi(\mathbf{r}) \right] = 0 , \qquad (2)$$

where $\ddot{\sigma}(\mathbf{r})$ is the spatially dependent conductivity tensor, written as

$$\ddot{\sigma}(\mathbf{r}) = \frac{\sigma(\mathbf{r})}{1+\left[\mu_e B_{probe}(\mathbf{r})\right]^2} \begin{bmatrix} 1 & \mu_e B_{probe}(\mathbf{r}) \\ -\mu_e B_{probe}(\mathbf{r}) & 1 \end{bmatrix}. \qquad (3)$$

In Eq. (3) $\sigma(\mathbf{r})=\mu_e n(\mathbf{r})e$ is the zero-field conductivity, with $\mu_e$ being the electron mobility, $n(\mathbf{r})$ the local electron density and $e$ the electron charge. The probe magnetic field $B_{probe}(\mathbf{r})$ is calculated by approximating the ML-MFM probe in both FM and A-FM states as a two point dipole system, where each dipole corresponds to the individual magnetic layer of the probe. The two dipole parameters (vertical and in-plane components of their magnetic moments and mean vertical distance from the probe apex $h_{mean}$) are provided by the RSTTF characterisation, introducing a variation in the range of ±5% to obtain the best fit with the experimental results. In the simulations, the in-plane component is oriented at 10° with respect to the vertical arm of the Hall cross (Fig. 5b), in order to take into account the in-plane projection of the probe during the experimental measurements. This component corresponds to the one labelled as $x$-axis component in the RSTTF characterisation (see Table 1). Moreover, in agreement with RSTTF analysis, the dipole positions are shifted in horizontal direction to take into account the presence of the separating Si layer, namely 40 nm for the thick ML-MFM probe and 25 nm for the thin one. To model probe oscillation effects, the simulated maps are obtained by subtracting the Hall voltage values obtained at the maximum probe-sensor distance ($h_{mean}$ + 44 nm) from the ones obtained at zero height of probe apex ($h_{mean}$ − 44 nm).

To reconstruct the non-negligible electrostatic effects found with the thin ML-MFM probe, a spatially dependent carrier density,

$$n(\mathbf{r}) = n_0 \left[1 - \psi(\mathbf{r})/E_F\right], \quad (4)$$

is introduced[38] describing in a phenomenological way the accumulation/depletion of charges in the graphene region underneath the probe, caused by the capacitive coupling with the probe. In Eq. (3), $E_F$ is the graphene Fermi energy, and $n_0$ is the electron density in the absence of electrostatic effects. The function,

$$\psi(\mathbf{r}) = \psi_0 \exp\left[-(\mathbf{r} - \mathbf{r}_0)^2 / d^2\right], \quad (5)$$

is the local potential profile induced by the probe, described as a Gaussian potential barrier with amplitude $\psi_0$ (corresponding to a maximum carrier density variation $\Delta n = n_0 \psi_0 / E_F$), centre position $\mathbf{r}_0$ in the sensor plane and characteristic length scale $d$. Parameters $\psi_0/E_F$ and $d$ are defined searching for the values that lead to the best fit with the experimental results reported in Figs. 6b and 6c.

**Acknowledgements**

The authors wish to thank D. Kurt Gaskill and Rachael L. Myers-Ward for providing the epitaxial graphene, Arseniy Lartsev for fabricating the Hall bar devices, and Sergey Vdovichev for developing the ML-MFM probes. This work was supported by the UK government's Department for Business, Energy and Industrial Strategy, EMRP/EMPIR under projects EXL04 (SpinCal) and S01 (NanoMag), EC grants Graphene Flagship (No. CNECT-ICT-604391) and NMS under the SC Graphene Project (No 119378). B. Gribkov acknowledges financial support by the CRDF grant #FSCX-14-61077-0, MES of Russia (Agreement 02.B.49.21.0003) and RFBR. L. A. Rodriguez and E. Snoeck acknowledge the "Conseil Regional Midi-Pyrénées", the European Union Seventh Framework Program under a contract for an Integrated Infrastructure Initiative Reference No. 312483-ESTEEM2 and the European FEDER for financial support within the CPER program and the French National Research Agency under the "Investissement d'Avenir" program reference No. ANR-10-EQPX-38-01.

**Author Contributions**

O.K., B.G, and V.N. designed the research, B.G. fabricated the ML-MFM probes, V.P. and H.C.-L. performed the MFM and SGM, B.G. performed the SEM, L.A.R., H.C.-L. and E.S performed the electron holography, V.N. and S.V. performed the qMFM and RSTTF and A.M. and E.Si. performed the numerical modelling. V.P., H.C.-L., B.G., L.A.R., A.M., V.N. and O.K. analysed the results and participated in writing the manuscript. V.P. prepared Figure 1, H.C.-L., L.A.R., and V.P. prepared Figure 2, V.N. prepared Figure 3-4, H.C.-L. and V.P. prepared Figure 5 and 6, and A.M., E.Si. and H.C.-L. prepared Figure 7. All authors reviewed the manuscript and discussed the results.

**Additional Information**

**Supplementary information** accompanies this paper at http://nature.com/srep

**Competing financial interests**: The authors declare no competing financial interests.